
\magnification=1200
\hsize=6truein\vsize=8.5truein

\font\bigbf=cmbx10 scaled\magstep1

\def\eg{{\it e.g. }}
\def\ie{{\it i.e. }}

\def\Tr{{\rm Tr}}
\def\mtilde{\widetilde{\bf m}}

\def\P{{\rm P}}


\def\avs#1{\langle{#1}\rangle}
\def\Epz#1#2#3#4{{Z\left|{#1\atop#3}{#2\atop#4}\right|}}
\def\dEpz#1#2#3#4{{Z'\left|{#1\atop#3}{#2\atop#4}\right|}}

\def\sumdash#1{{\mathop{{\sum}'}_{#1}}}
\def\sup#1{${#1}$}

\def\reff#1{\smallskip\par\noindent{${#1}$}}

\def\al{\alpha}
\def\be{\beta}
\def\ze{\zeta}
\def\La{\Lambda}
\def\om{\omega}
\def\itom{{\mit\Omega}}

\def\sect{{\vskip 10truept\noindent}}
\rightline {MUTP 92/1}
\vglue 1truein
\centerline{\bigbf A FINITE MODEL OF TWO-DIMENSIONAL}
\vskip 5truept
\centerline{\bigbf IDEAL HYDRODYNAMICS}
\vskip 15truept
\centerline{J.S.Dowker and A.Wolski}
\vskip 10truept
\centerline{\it Department of Theoretical Physics}
\centerline{\it The University,Manchester,England}
\vglue .75truein
\centerline{ABSTRACT}
\vskip 10truept
A finite-dimensional su($N$) Lie algebra equation is discussed that in the
infinite $N$ limit tends to the
two-dimensional, inviscid vorticity equation on the torus. The equation is
numerically integrated, for various values of $N$, and the time evolution
of an (interpolated) stream function is compared with that obtained from a
simple mode truncation of the continuum equation. The time averaged vorticity
moments and correlation functions are compared with canonical ensemble
averages.

\rightline {June 1992}
\vfill\eject
\sect {\bf 1.INTRODUCTION}

\noindent The vorticity equation for an ideal fluid on a two-dimensional
manifold ${\cal M}$ is
$${\dot{\zeta}}+\{\zeta,\psi\}=0,\eqno(1)$$where $\zeta$ is the vorticity and
$\psi$ the stream function related to $\zeta$ by
$$\zeta=-\Delta_2\psi.\eqno(2)$$ $\Delta_2$ is the Laplace operator on
${\cal M}$ and $\{f,g\}$ is the Poisson bracket of $f$ and $g$.

This equation has, of course, been the subject of numerous studies over the
years. It will be enough to mention the analysis of atmospheric motion
(in the zero height approximation) and the theory of turbulence.

A standard approach is to expand $\psi$ in modes of $\Delta_2$ so that (1)
becomes a coupled mode equation, the coupling coefficients being the structure
constants of the Poisson algebra, with factors involving the eigenvalues.

Precisely, define modes, $Y_\alpha$, and eigenvalues, $\lambda_\alpha$, by
$$\Delta_2 Y_\alpha=-\lambda_\alpha Y_\alpha$$ and expand $\zeta$
and $\psi$, $$\zeta=\sum_\alpha\zeta\alpha Y_\alpha=\sum_\alpha\lambda_
\alpha\psi\alpha Y_\alpha,\quad\quad\psi=\sum_\alpha\psi\alpha Y_\alpha
\eqno(3)$$ so that (1) reads
$$\lambda_\alpha\dot{\psi}\alpha+\lambda_\beta{C\alpha}_{\beta\gamma}
\psi\beta\psi\kappa=0,\eqno(4)$$where the structure coefficients are
defined by $$\{Y_\beta,Y_\gamma\}=Y_\alpha {C\alpha}_{\beta\gamma}$$ and
are given, by orthogonality, as an integral over three harmonics.
On the two-sphere, Elsasser\sup1 appears to have been the first to write down
this integral, although he does not refer to the Poisson algebra.
We will not enter into a detailed history of these sphere coefficients. They
occur in the work of Silberman\sup2 and of Baer and Platzman\sup3 in early
studies of
atmospheric vorticity. It was noted\sup{4,5}, sometime later, that the
coefficients were proportional to Clebsch-Gordan coefficients although
the calculation of the reduced matrix element was cumbersome.

The two-torus, T$2$, presents, in some aspects, a simpler situation and
its Poisson algebra was
first discussed by Lorenz\sup6 who was concerned to truncate an infinite
coupled
mode system (in the atmosphere) to the simplest nontrivial, finite one. The
same algebra was later investigated by Arnold\sup7, also in connection with
hydrodynamics.

The modes on the torus are plane waves, $\exp(i{\bf n.r})$, ($-\pi<x\le\pi,
-\pi<y\le\pi$)
and the eigenvalues are $\lambda_{\bf n}={\bf n}2$, ${\bf n}\in {\rm Z}2$.
The expansions of the stream function and vorticity are
$$\psi({\bf r})=\sum_{\bf n}\psi{\bf n}e{i{\bf n.r}},\quad
\zeta({\bf r})=\sum_{\bf n}\zeta{\bf n}e{i{\bf n.r}}\eqno(5)$$with
$\zeta{\bf n}=\lambda_{\bf n}\psi{\bf n}$. The structure constants are
$${C{\bf n}}_{\bf n',n''}={\bf n'}\wedge{\bf n''}\delta{\bf n}_{\bf n'+n''}.
\eqno(6)$$

The problem of mode truncation is a vital one in numerical weather
prediction and there seem to be no theoretical criteria for its optimum
solution. One point is that any truncation does
violence to the infinite set of conserved quantities (which may be taken to
be the integrated powers of the vorticity) for equation (1).
In Lorenz's truncation, for example, only the energy and the enstrophy were
conserved, and this was considered to be remarkable.

It is therefore of some interest to develop finite-mode approximations that
preserve more conserved quantities. Such models are suggested by
the fact\sup{8,9} that the Poisson algebra structure constants (or,
equivalently, the structure constants of the area-preserving diffeomorphism
group, SDiff$({\cal M})$), are the limits of the structure constants of SU$(N)$
as $N$ tends to infinity, after a simple change of normalisation.
We can say that the commutator
of two elements of the Lie algebra of SU$(N)$ ``corresponds", in the limit,
to the Poisson bracket of two functions on ${\cal M}$. It can be
seen that there are at least $N$ constants of the motion, corresponding to
the energy and the $N-1$ Casimir operators.

Zeitlin\sup{10} has also suggested and investigated these models in works which
appeared after our analysis was undertaken. There are certain differences of
detail and emphasis. A reference to finite models is also made by
Zakharov\sup{11}.
\sect{\bf 2. MATRIX ANALOGUE OF THE VORTICITY EQUATION ON T$2$}

\noindent Since we are interested in SU$(N)$ we first present some standard
algebraic material regarding its generators\sup{{12},8,9} which
we write in the Weyl\sup{13} form (see also Schwinger\sup{14})
$$J_{\bf n}=\omega{n_1n_2/2}g{n_1}h{n_2}\eqno(7)$$ where the unitary $N
\times N$ matrices $h$ and $g$ satisfy $hg=\omega gh$ and $gN={\bf 1}=hN$.
We choose $N$ odd and $\omega=\exp(2ik\pi/N)$ where $k$ and $N$ are coprime.
The periodicities,
$$J_{{\bf n}+N{\bf p}}=(-1){{\bf p}\wedge{\bf n}+p_1p_2}J_{\bf n},
\quad k\,\,{\rm odd}$$ and
$$J_{{\bf n}+N{\bf p}}=J_{\bf n},\quad k\,\,{\rm even},\eqno(8)$$
can be used to bring any ${\bf n}$ onto the $N\times N$ lattice, ${\cal C}_N$
(the unit cell), defined by $-(N-1)/2\le n_i\le(N-1)/2$. We occasionally use
${\cal C_\infty}$ to denote the entire square lattice, ${\rm Z}2$.

The most popular choice appears to be $\omega=\exp(4i\pi/N)$ because of the
simple periodicity (8).

The $J_{\bf n}$ satisfy the relations
$$J_{\bf n}\dagger=J_{-{\bf n}}=(J_{\bf n}){-1}$$
$$J_{\bf n'}J_{\bf n''}=\omega{{\bf n''}\wedge{\bf n'}/2}J_{\bf n'+n''}
\eqno(9)$$
$${1\over N} \Tr(J_{\bf n'}\dagger J_{\bf n''}{})=\delta{\bf n'}_{\bf n''},
\quad({\bf n',n''}\in{\cal C}_N).$$

Splitting up (9) gives the commutation and anticommutation rules,
$$[J_{\bf n'},J_{\bf n''}]=i{2\over k}\sin\left({k\pi\over N}{\bf n''}\wedge
{\bf n'}\right)J_{\bf n'+n''}\eqno(10)$$and
$$[J_{\bf n'},J_{\bf n''}]_+={2\over k}\cos\left({k\pi\over N}{\bf n''}\wedge
{\bf n'}\right)J_{\bf n'+n''}.\eqno(11)$$
Another way of writing (9) is
$$J_{{\bf n}_1}J_{{\bf n}_2}=\sum_{{\bf n}_3}e{-i{2\pi k\over N}A_{123}}
J_{{\bf n}_3}\,\delta_{{\bf n}_1+{\bf n}_2}{{\bf n}_3},\eqno(12)$$
where $A_{123}$ is the area of the triangle formed by the vectors
${\bf n}_1,{\bf n}_2,-({\bf n}_1+{\bf n}_2)$.

If ${\bf n}$ includes the origin, with $J_{\bf 0}={\bf 1}$, the
$J_{\bf n},\,{\bf n}\in{\cal C}_N$, form a complete operator basis,
$${1\over N}\sum_{\bf n}{J_{\bf n}}\dagger T {J_{\bf n}}{}={\bf 1}\Tr\,T.$$
(Schwinger\sup{14}. Incidentally Schwinger chooses $k=1$.)

We note\sup{8,9} that, as $N\rightarrow\infty$, the structure
constants in (10), for finite ${\bf n'}$ and ${\bf n''}$, tend to those of the
torus Poisson algebra, up to a normalisation constant.

Let $v$ and $w$ be two, time-dependent elements of su$(N)$.
They can be expanded in the Lie algebra generators $J_{\bf n}$,
$$v(t)=\sum_{\bf n}v{\bf n}(t)J_{\bf n},\quad w(t)=\sum_{\bf n}w{\bf n}(t)
J_{\bf n}.\eqno(13)$$ The summation over ${\bf n}$ is restricted to the
lattice ${\cal C}_N-\{{\bf O}\}$. If we wish to extend the summation to
${\cal C}_N$, as we do, $v{\bf 0}$ is set equal to zero for traceless $v$.
Hermiticity is equivalent to the conditions
on the coefficients$${v{\bf n}}*=v{-\bf n},\quad\quad {w{\bf n}}*=w{-
\bf n}.\eqno(14)$$

Consider the equation $${\dot v}+iN\beta[v,w]=0\eqno(15)$$
which we wish to compare to the hydrodynamic equation (1) with $v$ the
vorticity and $w$ the stream matrices.
$\beta$ is a constant that will be specified later. There is no real
signficance to its value since the overall normalisation is actually arbitrary
and could be absorbed into a redefined time.

In order to correspond with (2), we require that, as $N\rightarrow\infty$,
$v{\bf n}\rightarrow\lambda_{\bf n}w{\bf n}$.
Assuming that this has been achieved, the statement about the structure
constants is that, as $N$ tends to infinity,
the equation that the coefficients $w{\bf n}$ satisfy tends to the same
equation that the coefficients of the expansion of $\psi$ in torus modes
satisfy. Then, in the limit, we might hope to identify $w{\bf n}$ with
$\psi{\bf n}$. This will be made more precise later.

Our intention is to look upon these finite-dimensional models as
playing the role of consistent Lorenz-type truncations although it is not
clear, {\it a priori}, whether they will prove to be of practical interest.

We turn first to the relation between $v$ and $w$, and
it is here that we differ from Zeitlin\sup{10}. He simply sets
$v{\bf n}$ equal to $\lambda_{\bf n}w{\bf n}$. We feel that the relation
should be expressible directly in terms of the Lie algebra elements
themselves and it is not clear whether this is true for Zeitlin's relation.

Looking at (2), we require the Lie algebra analogue of the Laplacian. To find
this we recall the significance of the operators $g$ and $h$ in (7) as
stepping operators in the quantum mechanics on the discretised
circle\sup{13,14,15}.

It is easy to verify that
$${\cal L}J_{\bf n}\equiv\left({N\over2\pi}\right)2
\big(\big[[h,J_{\bf n}],h{-1}\big]+\big[[g,J_{\bf n}],g{-1}\big]\big)=
-\Lambda_{\bf n} J_{\bf n}$$where
$$\Lambda_{\bf n}(k)=\left({N\over\pi}\right)2\left(\sin2\big({k\pi n_1
\over N}\big)+\sin2\big({k\pi n_2\over N}\big)\right)\eqno(16)$$which we
recognise as proportional to the eigenvalues of the difference Laplacian on
the discretised torus $(2k\pi/N){\rm Z}_N\!\otimes\!(2k\pi/N){\rm Z}_N$.
The normalisation factor is chosen to give the correct continuum limit. If
$k=1$, $\Lambda_{\bf n}\rightarrow\lambda_{\bf n}={\bf n}2$ as $N\rightarrow
\infty$ for fixed ${\bf n}$. As a set, the $\Lambda_{\bf n}$
are independent of $k$. In fact $\Lambda_{\bf n}(1)=\Lambda_{{\bf n}_\P}(k)$
where the components of ${\bf n}_{{}_\P}$ are cyclic permutations of those of
${\bf n}$ according to
$${\bf n}_{{}_\P}=\bar k{\bf n}\,\,{\rm mod}\,(N,N),\eqno(17)$$where $\bar k$
is the mod inverse of $k$, \ie  $k\bar k=1$ mod $N$.
(See \eg\v C\' i\v zek\sup{16}.)

Another way of expressing this is to say that the discrete $\zeta$-function,
$$\sumdash{{\bf n}\in{\cal C}_N}{\exp(2ik\pi{\bf n.m}/N)\over\big
(\Lambda_{\bf n}(k)\big)s},$$ is independent
of $k$. The prime means that the ${\bf n=0}$ term is to be omitted. We note
that $J_{\bf 0}$, the unit matrix, is the zero mode, ${\cal L}J_{\bf 0}=0$.

It might be helpful to remark that the continuum $\nabla2$ can be written
in terms of repeated Poisson brackets,
$$\nabla2\psi=\left\{\{e{ix},\psi\},e{-ix}\right\}+
\left\{\{e{iy},\psi\},e{-iy}\right\}.$$

We take the operator ${\cal L}$ to be the discrete Lie algebra
analogue of $\Delta_2=\nabla2$ so that the generators $J_{\bf n}$ are the
analogues of the modes $Y_\alpha$, (plane waves on T$2$) as befits a
complete set. The relation
between $v$ and $w$ is thus written neatly as $v={\cal L}w$ and (15) becomes
$${\cal L}\dot w+iN\beta[{\cal L}w,w]=0.\eqno(18)$$

At this point it is convenient to discuss the conservation properties
of (5). We first need the fact that ${\cal L}$ is hermitian, \ie
$$\Tr\left(a\dagger {\cal L}b\right)=\Tr\left({\cal L}(a\dagger b\right)$$
where $a$ and $b$ are elements of su$(N)$. (Of course $a\dagger=a$.)
The trace is the finite analogue of integration over ${\cal M}$.

It is then easy to show that the quantity
$$E\equiv {1\over2N}\Tr(vw)={1\over2}\sum_{\bf n}\Lambda_{\bf n}w{\bf n}
w{-\bf n},\eqno(19)$$which we refer to as the energy, is time-independent.

Also, quite trivially and independently of the relation between $v$ and $w$,
the traces of powers of $v$ are conserved
$$S_l={1\over N}\Tr(vl)=\sum_{{\bf n}_1\ldots{\bf n}_l}v{{\bf n}_1}\ldots
v{{\bf n}_l}\cos\left({2k\pi\over N}A_{1\ldots l}\right),\eqno(20)$$
where $A_{1\ldots l}$ is the area of the $(l+1)$-gon in ${\cal C}_\infty$
with edges
${\bf n}_1,\ldots,{\bf n}_l,{\bf n}_{l+1}$ subject to the restriction
${\bf n}_{l+1}\equiv-\sum_{i=1}l{\bf n}_i=(0\,{\rm mod}\,N,0\,{\rm mod}\,N)$.

If $k$ is even the periodicity of the cosine allows one
to replace $A_{1\ldots l}$ by the area of the $l$-gon, ${\bf n}_1,\ldots,-\sum_
1{l-1}{\bf n}_i$. (In fact the whole polygon can be pulled back to fit
into the unit cell.)

There are $N-1$ independent $S_l$, $l=1,\ldots,N-1$, corresponding to the
anticommutator (11) \ie to the
Casimir invariants constructed from the symmetric $di_{jk}$ SU($N$) invariant
tensors\sup{17,18,19}. $S_1$ always vanishes. $S_2$ is the enstrophy, $\itom$.

These invariants also arise in the analyses of the generalised Euler
equations of rigid body motion\sup{9,20,21,22} except that
the group there is taken to be SO$(N)$ so that only the even powers remain.

As $N$ tends to infinity, (20) should become the continuum expression,
assuming that the $v{\bf n}$ tend to the $\zeta{\bf n}$ of (5).

The dynamical equations for the stream element coefficients are
$$\dot v{\bf n}(t)=\Lambda_{\bf n}{\dot{w}}{\bf n}(t)=-\sum_{\bf n',n''}
G{\bf n}_{\bf n',n''}\Lambda_{\bf n'}w{\bf n'}(t)w{\bf n''}(t)\eqno(21)$$
where the summations are restricted to lie on the lattice
${\cal C}_N-\{{\bf O}\}$ and where the coupling coefficients are given by
$$G{\bf n}_{\bf n',n''}=2N\beta\sin\left({k\pi\over N}{\bf n'}\wedge{\bf
n''}\right)\bar\delta{\bf n}_{\bf n'+n''}.\eqno(22)$$
The periodicity (8) has been incorporated by defining the (quasi-)periodic
delta, $\bar\delta$ with
$$\bar\delta{\bf n}_{\bf l}=\sum_{{\bf p}\in{\cal C}_\infty}\delta
{{\bf n}+N{\bf p}}_{\bf l},\quad k\,{\rm even}$$and
$$\bar\delta{\bf n}_{\bf l}=\sum_{{\bf p}\in{\cal C}_\infty}(-1){{\bf n}
\wedge{\bf p}+p_1p_2}\delta{{\bf n}+N{\bf p}}_{\bf l},\quad k\,{\rm odd}
\eqno(23)$$so that ${\bf n}$ can be restricted to the unit cell.
In (23), ${\bf l=n'+n''}$ and the sums are actually restricted to
${\bf p}\in{\cal C}_3$ because adding two elements of ${\cal C}_N$ can take
us only to the ``nearest neighbour" unit cells.

As $N\!\rightarrow\!\infty$, the $G{\bf n}_{\bf n',n''}$ tend to the Poisson
algebra structure constants (for fixed ${\bf n'}$ and ${\bf n''}$) and we
expect (21) to turn into Lorenz's torus equation. However
it is necessary to be careful when taking the $N\rightarrow\infty$ limit. One
cannot simply substitute the limiting form of (22) directly into
(21) because of the behaviour of terms for which $k{\bf n'}$ is of order
$N$, for example. To elucidate this limit we shall rewrite the equation in
coordinate representation but first, another motivation for this particular
step will be given.

The aim is to solve equation (21) numerically for given initial
conditions, and then to compare with the corresponding discussion of Lorenz,
\ie  with a simple truncation. Hence there arises a question concerning
the appropriate
quantity to construct once the coefficients have been computed. It is
possible to compare the coefficients directly but this is sensible only
for $N$ large, so making the comparison impractical. A stream ``function" is
needed for a global picture. That is, a continuous quantity constructed
from $w(t)$, for any finite $N$, that can be compared with the conventional
stream function $\psi({\bf r},t)$ after a numerical integration.

It has been noted\sup{23,24} that the limit $N\longrightarrow\infty$
is akin to the transition from quantum to classical mechanics with, in these
references, $4\pi/N$ playing the role of Planck's constant. This suggests
that we regard
(18) as a Heisenberg equation of motion and derive the corresponding classical
equation in the standard fashion using, say, coherent states\sup{25}.
This would give a concrete connection between $w$ and the ``classical"
stream function and will be pursued elsewhere.

We have not seen a discussion of coherent states in Weyl's finite
formulation of quantum mechanics although there are several applications
of the Wigner phase-space technique\sup{26,27,28}
to which we now turn.

This more formal point of view is provided by the representation of quantum
mechanics (called ``treacherous" by Groenewold\sup{29}) introduced by
Groenewold\sup{30},
Moyal\sup{31} and others, based on the Wigner\sup{32} phase-space distribution,
and much studied since. The quantum equations are replaced, exactly, by
a classical looking equation but with the
Moyal bracket (actually due to Groenewold) instead of the Poisson bracket.

In this approach, which, {\it a priori}, is distinct from the
coherent state method, one constructs the Weyl-Wigner
distributions, $\Tr(aJ_{\bf n})$, which are then interpreted as the Fourier
components of the classical quantity corresponding to the operator, $a$.
Usually one starts
from a classical quantity and then asks for the corresponding quantum
operator. This is the well known ordering problem.

A more general ordering\sup{30,33,34} is provided by
setting $$a({\bf m})={1\over N}\sum_{\bf n}\Tr\left(aJ\dagger_{\bf n}\right)
\Omega({\bf n})
e{i{2\pi\over N}k{\bf n.}\mtilde}\eqno(24)$$(usually $k=1$). The Weyl
ordering corresponds to
$\Omega=1$ and then $a({\bf m})$ is called the Wigner function.


If $a({\bf r})$ is to be real when $a$ is hermitian, the function $\Omega$
must satisfy $\Omega*({\bf n})=\Omega(-{\bf n})$ and we also want
$\Omega\rightarrow1$ as
$N\rightarrow\infty$. Typically, $\Omega({\bf n})$ is a trigonometric function
of the product $n_1n_2$.
A gaussian form for $\Omega$ is associated with normal or antinormal ordering.

The quantity that corresponds to the commutator $[a,b]$ is the Moyal bracket
(if $\Omega=1$) and, in this case, the coordinate-space representation of
the vorticity equation (15) is (the proof is given shortly)
$$\dot v({\bf m},t)+\{v,w\}_{\!{}_M}({\bf m},t)=0,\eqno(25)$$
where the {\it finite} Moyal bracket, $\{a,b\}_{\!{}_M}$ is defined by
$$\{a,b\}_{\!{}_M}({\bf m})={k3\over N\pi}\sum_{\bf m',m''}a({\bf m'})
b({\bf m''})\sin\left(\!{8k\pi\over N}  A\right),\eqno(26)$$with
$$  A={1\over2}({\bf m}\wedge{\bf m'}+{\bf m'}\wedge{\bf m''}+
{\bf m''}\wedge{\bf m})$$ the area of the dual triangle with vertices at
${\bf m},\,{\bf m'}$ and ${\bf m''}$.

It is clearly possible\sup{35} to generalise the Moyal bracket to allow for
the more general
ordering (24) involving $\Omega$ but we will not pursue this
point here except to say that (7) and (24) show that the different choices
for $k$ are related to the ordering question.

As mentioned before, one reason for introducing the coordinate-space
representation is that the infinite $N$ limit appears to be more transparent
than in the mode representation (21) which always remains discrete. We
will deduce a value for the constant $\beta$.

Equations (25) and (26) are now derived.
The finite Fourier relation we require reads
$${1\over N2}\sum_{{\bf n}\in{\cal C}_N}e{i{2\pi\over N}k{\bf n.m}}
=\sum_{{\bf p}\in{\cal C}_\infty}\delta{\bf m}_{N{\bf p}}\eqno(27)$$ and
the transform is defined by
$$a({\bf m})=\sum_{{\bf n}\in{\cal C}_N}a{\bf n}e{i{2\pi\over N}k{\bf n}.
\mtilde}
=\sum_{{\bf n}\in{\cal C}_N}a{{\bf n}_\P}e{i{2\pi\over N}{\bf n}.\mtilde},$$
$$a{\bf n}={1\over N2}\sum_{{\bf m}\in{\cal C}*_N}a({\bf m})e{-i{2\pi\over
N}k{\bf n}.\mtilde}.\eqno(28)$$
${\cal C}*_N$ is the dual lattice. We often identify ${\cal C}_N$ and
${\cal C}*_N$.

The expressions provide periodic extensions of $a({\bf m})$ and $a{\bf n}$
off the corresponding unit cells.

As a technical point of some interest, the appearance of the factor of $k$ in
(28) is related to the use of the eigenvalues $\La_{\bf n}(k)$ of (16). If we
had simply chosen to set $k=1$ in (16) (but {\it not} in the definition of
$\om$), then the $k$ in (28) must be unity too. There is nothing wrong in this,
but it would not then be possible to write the dynamical equations in purely
Lie algebra terms, as we have done in (18). We believe this is of more than
aesthetic importance.

As $N\rightarrow\infty$, with ${\bf r}=(2\pi/N)\mtilde$ and $a({\bf m})
\rightarrow a({\bf r})$, the formulae (28) turn into a standard Fourier series,
$$a({\bf r})=\sum_{\bf n}\bar a{\bf n}e{i{\bf n.r}},$$
$$\bar a{\bf n}={1\over(2\pi)2}\int_{T2} d{\bf r}\,a({\bf r})\,e{-i{\bf
n.r}}.\eqno(29)$$
and ${\bar a}{\bf n}=a{\bar{\bf n}}$ where $\bar{\bf n}$ stands for a pair
of reordered sets of {\it all} the integers. We have used the continuum
replacement
$$\sum_{{\bf m}\in{\cal C}*_N}\longrightarrow\left({N\over2\pi}\right)2
\int_{T2}d{\bf r}.$$

It is formally attractive to define the transformed generators, $J({\bf m})$,
by $$J({\bf m})=\sum_{{\bf n}\in{\cal C}_N}J_{\bf n}e{-i{2\pi\over N}k{\bf n}.
\mtilde},$$
$$J_{\bf n}={1\over N2}\sum_{{\bf m}\in{\cal C}*_N}J({\bf m})e{i{2\pi\over
N}k{\bf n}.\mtilde},\eqno(30)$$so that
$$v=\sum_{\bf n}v{\bf n}J_{\bf n}={1\over N2}\sum_{\bf m}v({\bf m})J({\bf m})
.$$ The $J({\bf m})$ are hermitian, and, as we have defined them, satisfy the
relations dual to (9)
$$\Tr J({\bf m})=N2,\, {\rm for\,all}\,{\bf m},$$
$$J({\bf m'})J({\bf m''})=\sum_{\bf m}f({\bf m'},{\bf m''};{\bf m})J({\bf m}),
\eqno(31)$$
$${1\over N2}\Tr\big(J({\bf m})J({\bf m'})\big)=N\bar\delta_{\bf m-m'},$$
$$J_{-\bf m}J({\bf m}')J_{\bf m}=J({\bf m'-m}).$$

These relations hold for all $k$ but an even value would be
preferred because of the implied simple periodicity of the $J_{\bf n}$.

If $k$ is even, a short calculation using (27) shows that the
composition constants are given by
$$f({\bf m'},{\bf m''};{\bf m})=\exp\left({i{8k\pi\over N} A({\bf m'},
{\bf m''},{\bf m})}\right).\eqno(32)$$ where $A({\bf m'},{\bf m''},
{\bf m}),=-A({\mtilde'},{\mtilde''},{\mtilde})$, is the
area of the triangle $({\bf m',m'',m})$ on the dual lattice, given before.

Consider the product of two operators (\ie  Lie algebra elements)
$$ab={1\over N4}\sum_{\bf m'}\sum_{\bf m''}
a({\bf m'})b({\bf m}'')J({\bf m'})J({\bf m''})\equiv{1\over N2}
\sum_{\bf m}{(a*b)}({\bf m})J({\bf m}),$$
all sums being over ${\cal C}_N$. This defines the $*$- or
Moyal-product. Therefore from (31)
$$(a*b)({\bf m})={1\over N2}\sum_{\bf m',m''}a({\bf m'})b({\bf m''})
f({\bf m'},{\bf m''};{\bf m}).\eqno(33)$$ Taking the commutator gives
$$[a,b]={1\over N2}\sum_{\bf m}[a,b]({\bf m})$$where
$$[a,b]({\bf m})=(a*b-b*a)({\bf m})={2i\over N2}\sum_{\bf m',m''}
a({\bf m'})b({\bf m''})\sin\left({8k\pi\over N}A({\bf m'},{\bf m''},
{\bf m})\right)$$ $$={2i\pi\over k3N}\{a,b\}_{\!{}_M}({\bf m}).\eqno(34)$$
Equation (15) can then be written as (25), with (26), as promised, if $\beta=
k3/2\pi$.

The continuum limit of (25) can be checked by replacing ${\bf m,m',m''}$ by
${\bf r,r',r''}$ respectively, where ${\bf r}=
(2\pi/N)\mtilde$ etc. (for all $k$). Then, in the infinite $N$ limit,
${\bf m}\rightarrow\infty$, the sums turn into integrals and, just as (28)
becomes (29), (26) goes over into $$\{a,b\}_{\!{}_M}({\bf m})\rightarrow$$
$$\lim_{N\rightarrow\infty}{k3N3\over4\pi5}\int\limits_{T2\times T2}\!
d{\bf r'}d{\bf r''}a({\bf r'})b({\bf r''})\sin\left({2kN\over\pi}  A
({\bf r,r',r''})\right)
=\{a,b\}({\bf r}),\eqno(35)$$the Poisson bracket, as required. k is assumed to
remain fixed and ${\bf r}\in{\rm T}2$. (The finite
integration ranges could be replaced by infinite ones, in the limit, to give
precisely the same integrals as in Baker\sup{36}.)
$  A({\bf r,r',r''})={1\over2}({\bf r}\wedge{\bf r'}+{\bf r'}
\wedge{\bf r''}+{\bf r''}\wedge
{\bf r})$ is the area of the coordinate-space triangle $({\bf r,r',r''})$.
The conclusion is that
equation (25) becomes the continuous Euler equation (1).

The coordinate-space representation also allows us to confirm the form of the
discrete Laplacian, ${\cal L}$. The standard expression for the
finite-difference Laplacian on {\bf T}$2$, scaled to give the correct
continuum
limit, is $$\Delta2J({\bf m})=\left({N\over2\pi}\right)2\sum_{<{\bf m}>}
\left[J(<\!{\bf m}\!>)-J({\bf m})\right],$$
where $<\!{\bf m}\!>$ are the nearest neighbours to ${\bf m}$. Thus
$$\Delta2J({\bf m})=\left(N\over2\pi\right)2
\big(J(m_1,m_2-1)+J(m_1,m_2+1)-2J(m_1,m_2)+$$
$$\quad\quad J(m_1+1,m_2)+J(m_1-1,m_2)-2J(m_1,m_2)\big).$$

It is easily
shown that $$\Delta2J({\bf m})=\left(N\over2\pi\right)2\left(
\big[[h,J({\bf m})],h{-1}\big]+\big[[g,J({\bf m})],g{-1}\big]\right).$$
Thus one can write$${\cal L}v={1\over N2}\sum_{{\bf m}\in{{\cal C}_N}*}
v({\bf m})\Delta2J({\bf m})={1\over N2}\sum_{{\bf m}\in{{\cal C}_N}*}
\Delta2v
({\bf m})J({\bf m})$$with $\Delta2v({\bf m})\rightarrow\nabla2v({\bf r})$,
as required.

The invariants $S_l$ too can be recast in terms of $v({\bf m})$. We
write $$S_l={1\over(N){2l}}\sum_{{\bf m}_1\ldots {\bf m}_l}v({\bf m}_1)
v({\bf m}_2)\ldots v({\bf m}_l)\,G({\bf m}_1,\ldots,{\bf m}_l).\eqno(36)$$
$G$ is related to a finite Fourier transform, (a Gaussian sum), by
$$G({\bf m}_1,\ldots,{\bf m}_l)={1\over N}{\rm Sym}\Tr\left(J({\bf m}_1)\ldots
J({\bf m}_l)\right),\eqno(37)$$ where from (30) and (12)$${1\over N}\Tr\left(J(
{\bf m}_1)\ldots J({\bf m}_l)\right)=\sum_{{\bf n}_1\ldots{\bf n}_l}e{-i
{2k\pi\over N}\left(\sum_1l{\bf n}_i.\mtilde_i-A_{1\ldots l}\right)}.$$

The symmetrisation on the ${\bf m}_i$ can be
performed in various ways. Simply reversing their order gives
$$G({\bf m}_1,\ldots,{\bf m}_l)=\sum_{{\bf n}_1\ldots{\bf n}_l}
e{-i{2k\pi\over N}\sum_1l{\bf n}_i.\mtilde_i}\cos\left({2k\pi\over N}
A_{1\ldots l}\right)\eqno(38)$$ with the mod $N$ condition on
$\sum{\bf n}_i$.

Another formula results on combining the $J({\bf m})$ in (37) using the
composition law (31) to give
$${1\over N}\Tr\left(J({\bf m}_1)\ldots J({\bf m}_l)\right)=$$ $$\sum_{
\{{\bf m}_{i'}\}}
f({\bf m}_1,{\bf m}_2,{\bf m}_{3'})f({\bf m}_{3'},{\bf m}_3,{\bf m}_4)\ldots
f({\bf m}_{{(l-1)}'},{\bf m}_{l-1},{\bf m}_l).\eqno(39)$$

It is interesting to check that, in the infinite $N$ limit, $S_l$ becomes
the integrated power of the vorticity function, that is, up to a normalisation
factor,
$$ S_l\longrightarrow\int_{T2} d{\bf r}\,v({\bf r})l.\eqno(40)$$
and it is instructive to carry the limit through completely in coordinate-
space {\it after} the summations over the ${\bf n}_i$ have been done.

The behaviour of the function $G({\bf m}_1,\ldots,{\bf m}_l)$ as $N$ tends to
infinity is required. As usual we set ${\bf r}_i=(2\pi/N)\mtilde_i$ and write
$G({\bf r}_1,\ldots,{\bf r}_l)$.
Taking the expression (36) for the invariant
$S_l$, rescaling the $\mtilde_i$ to the ${\bf r}_i$ and changing the
summations into integrations produces
$$S_l\longrightarrow{1\over(2\pi){2l}}\int_{T2\times\ldots T2}d
{\bf r}_1\ldots d{\bf r}_l\,
v({\bf r}_1)\ldots v({\bf r}_l)\,G({\bf r}_1,\ldots,{\bf r}_l).\eqno(41)$$

The simplest nontrivial example is $l$ equal to three, when the polygons are
triangles. Then, immediately, from (31) and (32)

$$G({\bf m}_1,{\bf m}_2,{\bf m}_3)=N2\cos\left({8k\pi\over N}
  A({\bf m}_1,{\bf m}_2,{\bf m}_3)\right),$$
On rescaling, $  A({\bf m}_1,{\bf m}_2,{\bf m}_3)$ becomes
$(N/2\pi)2  A({\bf r}_1,{\bf r}_2,{\bf r}_3)$ and
we can now apply another formula given in Baker\sup{36},
$$\lim_{N\rightarrow\infty}\left({kN\over\pi}\right){2\nu}\!\!\!\int_
{-\infty}\infty d{\bf r}_1\ldots
d{\bf r}_{2\nu+1}\cos\bigg(\!{2kN\over\pi}\big(  A({\bf r}_1,
{\bf r}_2,{\bf r}_3)+  A({\bf r}_1,{\bf r}_4,{\bf r}_5)+\dots$$
$$\ldots+  A({\bf r}_1,{\bf r}_{2\nu},{\bf r}_{2\nu+1})\big)\bigg)
F({\bf r}_1,{\bf r}_2\ldots{\bf r}_{2\nu+1})$$ $$\quad\quad\quad=
\int_{-\infty}\infty d{\bf r}_1 F({\bf r}_1,\ldots,{\bf r}_1),\eqno(42)$$ for
$\nu=1$ to
give the desired continuum integral of $v({\bf r})3$ in the infinite $N$
limit. We note that in (42), $F$ must be a reasonable function.

The case of any $l$ will now be discussed. Although (39) is completely in
coordinate space it is not in a convenient form for the application of (42).
In fact, the general form of $G({\bf m}_1,\ldots,{\bf m}_l)$ can be given
with no summations. The expression depends on whether $l$ is even or odd and,
in fact, we shall restrict the discussion to odd $l$ for brevity.

We start from the form (38) and
begin by replacing ${\bf n}_l$ by $\bar{\bf n}_l\equiv-\sum_1{l-1}{\bf n}_i$.
This is allowed because of the periodicity of the exponential and the cosine.
We might then just as well rename $\bar{\bf n}_l$ as ${\bf n}_l$ and
restrict the sum (38) to {\it closed} $l$-gons, as mentioned earlier.

The evaluation proceeds by alternate
application of (27) and imposition of the resulting $\bar\delta$. The choice
of which ${\bf n}_i$ to sum over is crucial to obtaining a simple,
symmetrical result. It is convenient to first perform a cyclic permutation
of the ${\bf n}_i$ (under which $A_{1\ldots l}$ is invariant) so that
${\bf n}_b$ becomes ${\bf n}_i$ where $b$ is the next integer after $l/4$.
Then, performing the sums in the order ${\bf n}_l$ downwards, we find
$$G({\bf m}_1,\ldots,{\bf m}_l)=N{l-1}\cos\left({8k\pi\over N}
E\big(\{{\bf m}_i\}\big)\right),\eqno(43)$$where $E\big(\{{\bf m}_i\}\big)$
is given by $$E\big(\{{\bf m}_i\}\big)
=\sum_{i=2,4,\dots l}  A({\bf m}_1,{\bf m}_i,{\bf m}_{i+1})+
\sum_{i\ne j\atop{{i=2,4,\dots}\atop{j=2,4,\ldots}}}{l-1}
({\bf m}_i-{\bf m}_{i+1})\wedge({\bf m}_j-{\bf m}_{j+1})$$
$$\equiv  A\big(\{{\bf m}_i\}\big)+B\big(\{{\bf m}_i\}\big).\eqno(44)$$
This is a closed form for the Gauss sum.
The first summation is the area of $(l-1)/2$ triangles connected at the
vertex ${\bf m}_1$ in a windmill sail pattern. The second sum is that of
the cross-products of all pairs of vane ends.

Equation (43) with (44) yields (41) with $G$ given by
$$G({\bf r}_1,\ldots,{\bf r}_l)=N{l-1}\cos\left({2kN\over\pi}\big(
A(\{{\bf r}_i\})+B(\{{\bf r}_i\})\big)\right)$$
$$=N{l-1}\left(\cos\big({2kNA\over\pi}\big)\cos
\big({2kNB\over\pi}\big)-\sin\big({2kNA\over\pi}\big)\sin\big({2kNB\over
\pi}\big)\right)\eqno(45)$$in terms of rescaled quantities.

A completely immediate application of (42) to (41) is not possible because
the function $F$ now contains $N$.
However we note that the effect of the $\cos\big(2kN  A/\pi\big)$
terms in (42) in the $N\rightarrow\infty$ limit is to force ${\bf r}_i$ to
equal ${\bf r}_{i+1},\,(i=2,4,...,2l)$ and, also, both to equal ${\bf r}_1$.
Since $B$ has a product structure (see (44)), the $\cos(2kNB/\pi)$ factor can
be
replaced by unity, in the limit. We also note that the same equation as (42),
but with a sine (as used in (35) with an extra factor of $N$) gives a
result of the order of $1/N$ and so the second term in (45) goes away.
Equation (42) can now be applied directly with $F({\bf r}_1,\ldots,{\bf r}_l)=
v({\bf r}_1)\ldots v({\bf r}_l)$ to give the required continuum expression
(40). Our discussion of the coordinate representation of the invariants ends at
this point.

We can now give a reasonable answer to a question posed earlier regarding the
appropriate quantity to construct from the matrix $w(t)$ that can be compared
with the continuum steam function, $\psi({\bf r},t)$, resulting from a
standard mode truncation of Euler's equations (1). The Fourier transform (28)
suggests the simple interpolation
$$w({\bf r},t)=\sum_{{\bf n}\in{\cal C}_N}w{{\bf n}_\P}(t)e{i{\bf
n.r}},\eqno(46)$$
as a possible analogue of $\psi({\bf r},t)$. For convenience, it is this
quantity that is plotted but it is clear of course that there cannot be a
unique quantity corresponding to $\psi$.

We note that the Fourier coefficients in (46) are evaluated at the permuted
points ${\bf n}_\P$. This means that, for any $k$, the plane-wave modes with
the smaller $|n_1|,|n_2|$ are associated with the smaller eigenvalues,
$\Lambda_{{\bf n}_\P}(k)=\Lambda_{\bf n}(1)$, as occurs in the
continuous case. A naive application
of the $N\rightarrow\infty$ limit to the momentum-space equations, (21) and
(22), does not give the correct result.

There is a peculiarity in that the coordinate-space treatment of the
$N\rightarrow\infty$ limit is not easily available to us for what appears to
be the simplest value of $k$ (from the Fourier transform point of view) namely
unity. For finite $N$, the models for different
values of the parameter $k$ seem to be distinct. Our treatment shows,
however, that they will all yield the same continuum limit but we can vouch
for our coefficients only when $k$ is even.

\sect{\bf 3. MAXIMUM SIMPLIFICATION}.

\noindent We now turn to the practical solution of equations (21), along
the lines of Lorenz's calculation\sup6.

Since $\psi$ is real its Fourier coefficients satisfy the
condition ${\psi{\bf n}}*=\psi{-\bf n}$. The obvious finite equivalent
is the hermiticity of $w$, (14).

Lorenz\sup{6} notices that if the
coefficients are chosen to be real at some initial time, they will remain
real throughout the time development. The reality condition means that
${w{\bf n}}=w{-\bf n}$ and using the symmetries
$\Lambda_{-\bf n}=\Lambda_{\bf n}$ and $ G{\bf n}_{\bf n',n''}=G{-\bf n}_
{\bf-n',-n''}=-G{\bf n}_{\bf n'',n'}$ we can deduce from (21) that
$$\Lambda_{\bf n}{d\over dt}(w{\bf n}-w{-\bf n})={1\over2}\sum_{\bf n',n''}
(\Lambda_{\bf n'}-\Lambda_{\bf n''})G{\bf n}_{\bf n',n''}(w{\bf n'}+
w{-\bf n'})(w{\bf n''}-w{-\bf n''})\eqno(47)$$ showing that if the condition
$w{\bf n}=w{-\bf n}$ is valid for all ${\bf n}$ at some time, it remains
valid.

For $N=3$ the number
of independent real coefficients is four. This is the smallest number that
we can take for a consistent group theoretical structure. Lorenz makes a
further identification, that is also propagated in time, and this reduces his
number to three. In general, the number of coefficients in our maximum
simplification is $(N2-1)/2$.

For real $w{\bf n}$, $w$ can be rearranged into
$$w={1\over2}\sum_{\bf n}w{\bf n}(J_{\bf n}+J_{-\bf n}) $$
The combination of generators that occurs on the right hand side gives just
the generators of the subgroup U(SU$\left((N+1)/2\right)\otimes
{\rm SU}\left((N-1)/2\right)$ which means that the torus has actually been
turned into a tetrahedron\sup{37} which might not be unreasonable
for discussing atmospheric motion on the whole earth.

In all our calculations  $k$ was set equal to two. The numerical procedure
consisted of choosing  an
initial distribution of the coefficients $w{\bf n}$ and then integrating (21)
by standard routines. The results for the stream ``function" were
displayed in coordinate space using the Fourier interpolation (46). For each
odd value of $N$ from 3 to 31 the results were compared with those for a
simple trunctation method using the same number of modes.

The conservation of $E$ and of the $S_l$ was used as a check of the algorithms
and algebra.

The initial configurations are discussed in the next section.
\vfill\eject
\sect{\bf 4.VORTICES ON A LATTICE}.

\noindent For the sake of having something definite, it is interesting to
propagate a system of lattice vortices. A suitable set of initial
stream coefficients for a single vortex situated at ${\bf m=m}_i$ would be
$$w_i({\bf m})=\sumdash{{\bf n}\in{\cal C}_N}\,{\exp\big(2ik\pi{\bf n.}
(\mtilde-\mtilde_i)/N\big)\over\Lambda_{\bf n}(k)}.\eqno(48)$$ They are
independent of $k$. Further, $\sum_{\bf m} w_i({\bf m})=0$. The mode
coefficients are $w{\bf n}_i=\exp\big(2ik\pi{\bf n.m}_i/N\big)/
\Lambda_{\bf n}(k)$ if ${\bf n}\ne{\bf 0}$ and $w{\bf 0}_i=0$. For the
vorticity, $v{\bf n}_i=\exp\big(2ik\pi{\bf n.m}_i/N\big)$ if
${\bf n}\ne{\bf 0}$ with $v{\bf 0}_i=0$. In coordinate space, $v_i({\bf m})=
N2\delta{\bf m}_{{\bf m}_i}-1$ showing that the vorticity is mostly
concentrated
at the point ${\bf m}_i$, justifying the term ``vortex". The smaller, negative
value of
$-1$ makes the total ``integrated" vorticity, $\sum_{\bf m}v_i({\bf m})$,
zero, as necessitated by the compactness of the domain. However it is
not possible to conserve the vorticity located (in some sense) at ${\bf m}_i$,
which can leak away.

As $N$ and ${\bf m}$ tend to infinity, the discrete stream function,
$w_i({\bf m})$ of (48),
turns into the Epstein $\zeta$-function, except at $x=0$, $y=0$,
$$w_i({\bf m})\rightarrow\sumdash{{\bf n}\in{\cal C}_\infty}\,{\exp\big(i
{\bf n.r}\big)\over{\bf n}2}=\Epz00{x/2\pi}{y/2\pi}(2),\eqno(49)$$in Epstein's
notation\sup{38}. For simplicity, we have set $k=1$ and ${\bf m}_i={\bf 0}$.
$x$ and $y$ are the coordinates of ${\bf r}$ with $x=-2\pi m_2/N$ and
$y=2\pi m_1/N$. In the limit we would regard $x$ and $y$ as being continuous
and
restricted to the range $-\pi$ to $\pi$. It should be remarked that for any
fixed value of ${\bf m}$, the difference between $w_i({\bf m})$ and the
$\zeta$-function evaluated at ${\bf r}=2\pi{\mtilde}/N$, will be a nonzero
constant that decreases with increasing ${\bf m}$.

We can now compare (49) with some results for the stream function on the torus
derived in earlier times. Greenhill\sup{39} and Hicks\sup{40} give the
stream function in a rectangle. The expression on the torus can be
found in the intermediate calculation, but perhaps the easiest method of
proceeding is the following.

The stream function on the torus for a vortex at the origin is given as
the image expression, (cf \sup{40}),
$$\psi(x,y)={\kappa\over4\pi}\sum_{\bf M}\ln{(x/2\pi+M_1)2+(y/2\pi+M_2)2
\over(M_12+M_22)},$$where the sums run over all the integers and $x$
and $y$ are restricted to the range $-\pi$ to $\pi$.
Up to an additional constant,
$$\psi(x,y)=-{\kappa\over2\pi}{\partial\over\partial s}\bigg(\sum_{\bf M}
{1\over\big((x/2\pi+M_1)2+(y/2\pi+M_2)2\big){s/2}}\bigg)_{s=0}=$$
$$=-{\kappa\over2\pi}{\partial\over\partial s}\bigg(\Epz{x/2\pi}
{y/2\pi}00(s)\bigg)_{s=0}\equiv-{\kappa\over2\pi}\dEpz{x/2\pi}{y/2\pi}00(0).$$
Evaluation of the transformation formula\sup{38}
$$\pi{-s/2}\Gamma(s/2)\Epz{h_1}{h_2}00(s)=
\pi{s/2-1}\Gamma(1-s/2)\Epz00{-h_1}{-h_2}(2-s)$$at $s=0$ yields
$$\dEpz{h_1}{h_2}00(0)={1\over2\pi}\Epz00{-h_1}{-h_2}(2).$$
since $$\Epz{h_1}{h_2}00(0)=0,$$ if $h_1$ and $h_2$ are not integers.
Hence $$\psi(x,y)=-{\kappa\over(2\pi)2}
\Epz00{x/2\pi}{y/2\pi}(2).\eqno(50)$$Looking at (49), it can be seen that
the discrete stream function $w_i({\bf m})$ tends to the torus stream function,
$\psi(x,y)$, as $N\rightarrow\infty$ for strength $\kappa=-4\pi2$.

Hicks\sup{40} gives an expression for $\psi(x,y)$ in terms of simpler
$\theta$-functions. This is, up to the usual additive constant,
$$\psi(x,y)={\kappa\over4\pi}\left[{1\over4\pi}(x2-y2)+\ln\big(\theta_1
(z/2\pi,i)\theta_1(z*/2\pi,i)\big)\right],\eqno(51)$$ where $z=x+iy$ and
$\theta_1(u)$ is denoted by $H(2Ku)$ in Hicks and Greenhill. Using the Jacobi
transformation formula for $\theta$-functions, it is easily checked that
$\psi(x,y)$ is symmetrical in $x$ and $y$.
Our definitions of $\theta$-functions are those of Oberhettinger and
Magnus\sup{41} where a brief discussion of the motion of vortex systems can
also be found in chapter 4.

Incidentally, Kronecker\sup{42} reduced the Epstein $\zeta$-function (49) to a
form in $\theta$-functions,
$$\Epz00{h_1}{h_2}(2)=2\pi2h_12-\pi\ln{\theta_1(u_1,\omega_1)\theta_1(u_2,-
\omega_2)\over\eta(\omega_1)\eta(-\omega_2)},\eqno(52)$$ where $\eta$ is
the Dedekind function,
$\eta(\omega)=q{1/12}\prod_1\infty(1-q{2n})$. The variables $h_1$ and
$h_2$ are to be taken in the range from $0$ to $1$.
For the square torus,
$\omega_1=\omega_2*=i$, $q=e{-\pi}$ and $u_1=u_2*=(x+iy)/2\pi$.
Comparing (52) with (50) and (51), we see that Hicks and Greenhill had earlier
obtained an equivalent reduction. Kronecker's formula has been rediscovered a
number of times since. These formulae can be used for numerical evaluation of
$\psi$, (50), although there exists a form in terms of incomplete
$\Gamma$-functions that is generally more efficient, except for small $x+iy$.

For a single vortex, a glance at equation (21) reveals that $\dot w{\bf n}$
vanishes for all ${\bf n}$ and the vortex is thus stationary. The energy is
given by $E=w(0)=\sum1/\Lambda_{\bf n}(k)$ and the enstrophy is $S_2=N2-1$.
Both these quantities diverge as $N\rightarrow\infty$.

The stream matrix coefficients for a set of vortices is
$$w({\bf m})=\sum_i{\kappa_i\over4\pi2}w_i({\bf m})$$ where the $\kappa_i$ are
the vortex strengths. For two, equal vortices, setting ${\bf m}_2= -{\bf m}_1$
and $\kappa_2=\kappa_1=-4\pi2$, we get real coefficients $w{\bf n}=2\cos
\big(2k\pi{\bf n.m}_1/N\big)/\Lambda_{\bf n}(k)$, $w{\bf 0}=0$ so that we can
place this configuration on the tetrahedron. The enstrophy equals $2(N2-2)$
while the next invariant is $4(16-N2)$.

The systems with $N$ from 3 to 31, were evolved in time. We present the
results in Figure 1 for SU(9) as being typical.

The initial position of the vortices  was at ${\bf m}_1=(1,0)$.
In our view no particular pattern emerges.

In any finite scheme, the localisability of individual vortices is lost,
the vorticity can become redistributed and it is difficult to model the
motion of ideal, point vortices in this way unless, possibly, $N$ is extremely
large. However the evolution for SU(31), shown in Fig.2, offers no evidence
for such a trend.

The results for SU(5) were somewhat atypical and are displayed in Fig.3. The
vortices appear to be rotating around each other.

The mode-truncation calculation starting from the same stream function yields
results of a generally similar nature.

A number of other initial configurations were also propagated with, again,
\break roughly comparable outcomes.

\sect{\bf 5. STATISTICAL BEHAVIOUR.}

\noindent The numerical results so far presented are for relatively short
time evolution. It has been suggested by Kraichnan\sup{43} that two-dimensional
turbulence can be statistically modelled on the assumption that the system
is ergodic and can, after a sufficiently long time be described by a
microcanonical or even a canonical distribution. In the latter case, two
``temperatures" can be introduced as Lagrange multipliers for the conserved
quadratic quantities, the energy $E$ and the enstrophy $\itom$.

A number of computer experiments have been performed in both the viscid and
inviscid cases on the truncated versions of the Euler and Navier-Stokes
equations to test this idea. The results are suggestive but not conclusive.

The models discussed in the present work allow a similar numerical analysis.
These systems, having more than just the two conserved quantities, $E$ and
$\itom$, of the truncated theory, might provide a more realistic arena in which
to explore the statistical hypothesis.

With this in mind the systems were evolved for long times, at a reduced
numerical accuracy for speed purposes. The quantities, $E$ and the
$S_l$ were found typically to be conserved to one part in $105$ to $106$ over
the extended time period.
Vorticity moments and correlation functions were evaluated by simple time
averaging since actual ensemble averaging was impractical. The results were
compared with the canonical distribution values, where appropriate.

The basic theory can be found in Kells and Orszag\sup{44}, for example, and
so we need not give many details.

The canonical vorticity distribution is given by
$$P(\ze)={1\over Z}\exp\big(-\sum_{\bf n}(\al\Lambda_{\bf
n}{-1}+\be)|{\ze{\bf n}}|2\big)$$
with the partition function, $Z=\int P(\ze)d\ze$. The relation with the
parameters in Ref 44 is $\al=C{-1}$ and $\be=D{-1}$.

Ensemble averages are $$\avs{F(\ze)}=\int F(\ze)P(\ze)d\ze.$$ In terms of the
two temperatures $\al$ and $\be$ the moments of the vorticity are
$$\avs{(\ze{\bf n})p}=
{(2p-1)!!\over 2p}\left(\al\Lambda_{\bf n}{-1}+\be\right){-p}\eqno(53)$$

The time-averaged moments are
$$M_p({\bf n},T)={1\over T}\int_0T\big|\ze{\bf n}\big|pdt,$$
which, for large $T$, are to be compared with the ensemble averages (53).

The mode correlation function $C(s)$ is defined by
$$C({\bf n},T;s)={1\over TM_2({\bf n},T-s)}\int_0{T-s}\ze{\bf n}(t)\ze{\bf
n}(t+s)dt.$$The prefactor is a normalisation.
If $C(s)$ does not tend to zero with increasing $s$ the system is probably
not ergodic and cannot be described by a statistical ensemble.

Three starting distributions were chosen. One was the double vortex discussed
in section 4, another was a vortex-antivortex pair and a third was a more or
less random arrangement. In the latter
case the coeffients in the truncated model were adjusted to give the same
energy and enstrophy as the corresponding SU($N$) model so that the
calculated canonical temperatures $\al$ and $\be$ should be the same.

Figs.4, and 5 display some results using the double vortex starting
configuration for SU(9). The figure captions are descriptive. Fig.6 shows
the long time evolution of the corresponding stream function. For the double
vortex the values of the two temperatures were calculated to be
$\al=2.9255188\times10{-2}$ and $\be=0.24223645$.

The presentation is limited to these data sets for reasons of space and
also because one would really like to explore much larger $N$ values. More
extensive data and discussion is contained in the Manchester PhD thesis
of A.W.

As a measure of the accuracy of the evolution algorithms, we tested the
conserved
quantities. The following are the values of the SU(9) energy and the first
three Casimirs, $S_2$, $S_3$ and $S_4$, at $t=0$ (the first number in the
brackets) and at $t=1000$ (the second number): $E(14.7549152,14.7549145)$,
$S_2(158.000037,158.000058)$, $S_3(-308.0002,-308.0035)$,
$S_4(38022.0096,38022.0630)$. The integer parts of the $S_l$ are the exact
values. The evaluation of the highest Casimirs from the stream function
coefficients is very time consuming due to the multiple summations.

The statistical results are inconclusive. We find no indication of
non-ergodic
behaviour but the evidence for a statistical description is still only
suggestive. The results are perhaps better than one would expect for
systems with a small number of modes when treated by a canonical distribution
which, moreover, ignores the other conserved quantities.
The microcanonical distribution would be more relevant but the evaluation
of the statistical averages is then itself an involved numerical procedure
which we have chosen to avoid.

The trunction results differ in no essential aspect from the corresponding
SU($N$) cases except that the vorticity second moment of the highest mode
does not relax to the canonical distribution value for large times.

There is also nothing in particular that distinguishes the other starting
configurations although our numerical experiments are not yet very
extensive in this respect.

The statistical mechanics of systems of this type, with many conserved
quantities, remains to be elaborated. Zeitlin\sup{10} makes some relevant
remarks on the structure of the phase space.
\vfill\eject

\sect\noindent{\bf 6. COMMENTS AND CONCLUSION}.

\noindent In this paper we have been mainly concerned to set up the SU($N$)
models and to discuss the nature of the infinite $N$ limit, which is not
obvious. We have also presented a sample selection of numerical results which
should be regarded as exploratory rather than definitive.

Our numerical approach was unsophisticated. For the truncated system
(which is sometimes referred to as the Galerkin approximation) some
impressive speed advantages could have been achieved by using
pseudo-spectral\sup{45} or collocation methods combined with fast Fourier
transforms. This is, apparently,
not possible in the SU($N$) models because of the nonlocal coupling terms in
the coordinate-space form (25). The whole point of the pseudo-spectral method
is to calculate the mode-coupling terms in the representation in which they
are local. For the SU($N$) cases we do not seem to have this option. Without
some technique corresponding to the pseudo-spectral one, the SU($N$) models
could never be viewed as numerical alternatives to the standard truncation or
finite element methods.

It might be expected that, as $N$ becomes bigger, the
results for the group model and those for the truncated system should
approach one another. There was no evidence for this in the short-time
evolutions that we have performed. Also there was no indication that the
quantities conserved in the continuum theory were progressively better
conserved
in the truncated versions as $N$ increased. Perhaps the values of $N$ are
still too small or it might be that the $N\rightarrow\infty$ limit has not
been closely enough considered and that the expectation is unfounded.

More disturbing is the oscillatory behaviour of an ``entropy" $\sum_{\bf n}
\log\ze_n/\La_{\bf n}$ as a function of time. These evaluations are at a
preliminary stage and
have not been displayed. They may indicate that the systems are not ergodic or
that the number of modes is small.

A corresponding analysis can be performed on the two-sphere.
Although the coupling coefficients are more complicated, the eigenvalues
are simpler, being the same as in the $N\rightarrow\infty$ limit. A
discussion will be presented elsewhere.

Calculations have also been done on a triangular lattice, corresponding
to a regularly slanted torus. For real mode coefficients, the results are
relevant for motion on the surface of a regular tetrahedron whereas the
square torus discussed in this paper gives a degenerately flat
tetrahedron. It would not model the Earth too well. This tetrahedron is, in
fact one of the flat Riemann surfaces discussed elsewhere\sup{37} and it would
be possible to extend the present calculation to these.

Whether or not this whole class of models proves to be of use in approximating
continuous theories, they at least provide an interesting set of dynamical
systems.

More realistically the effect of viscosity could be investigated by
analysing the Navier-Stokes equation.

Another interesting question concerns the Lagrangian stability of the motion
\ie of the trajectories in the Lie {\it group}. It is known that the Eulerian
(or Lie algebra) motion can be stable yet the Lagrangian one unstable in the
continuous case, being related to the sectional curvature of the group of
area-preserving diffeomorphisms. It is amusing to consider the discrete
analogue of this.

The computations were carried out on a Hewlett-Packard workstation.
Transference to a
more powerful machine is planned and it is hoped to reach large values of N.

The recent preprint by Rankin\sup{46} contains some information on the
SU($\infty$) limit, mostly in a particle physics context.

After this work was completed we were apprised of the paper by Miller et al
\sup{47} in which this finite class of models is also discussed. No numerical
calculations are given and there are a number of formal differences
with our setup particularly the choice of the discrete Laplacian.
\vfill\eject
\sect {\bf    REFERENCES.}
\reff{1} Elsasser,W.M. 1946 Phys.Rev.{\bf 69} 106.
\reff{2} Silberman,L. 1954 J.Met.{\bf 11} 27.
\reff{3} Baer,F. and Platzman,G.W. 1961 J.Met.{\bf 18} 393.
\reff{4} Jones,M.N. 1970 Planet.Space Sci.{\bf 18} 1393.
\reff{5} Thiebaux,M.L. 1971 J.Atmos.Sci.{\bf 28} 1294.
\reff{6} Lorenz,E.N. 1960 Tellus {\bf 12} 243.
\reff{7} Arnol'd,V. 1966 Ann.Inst.Fourier {\bf 16} 319.
\reff{8} Hoppe,J. 1986 ``Quantum theory of a relativistic surface", in ``
Workshop on Constraints theory and relativistic dynamics", edited by
Longhi,G. and Lusanna,L. (World Scientific, Singapore).
\reff{9} Fairlie,D.B., Fletcher,P. and Zachos,C.K, 1989 Phys.Letts.{\bf B218}
203
; 1990 J. Math. Phys. {\bf31},1088; Fairlie,D.B. and Zachos,C.K. 1989
Phys.Letts. {\bf B224} 101.
\reff{10} Zeitlin,V. 1991 Physica {\bf D49} 353.
\reff{11} Zakharov,V.E. 1989 Funct.Anal.Appl.{\bf 23} 24.
\reff{12} 't Hooft,G. 1981 Comm.Math.Phys. {\bf 81} 267.
\reff{13} Weyl,H. 1931 ``Theory of groups and quantum mechanics" (Methuen,
London).
\reff{14} Schwinger,J. 1960 Proc.Nat.Acad.Sci.{\bf 46} 570.
\reff{15} Floratos,E.G. 1989 Phys.Lett.{\bf B233} 395.
\reff{16} \v C\' i\v zek,V. 1986 ``Discrete Fourier transforms and their
applications" (Adam Hilger, Bristol).
\reff{17} Biedenharn,L.C. 1963 J.Math.Phys.{\bf 4} 436.
\reff{18} Klein,A. 1963 J.Math.Phys.{\bf 4} 1283.
\reff{19} Sudbery,A. 1990 J.Phys.{\bf A23} L705.
\reff{20} Weyl,H. 1922 ``Space, time, matter"(Methuen,London).
\reff{21} Manakov,S.V. 1976 Func.Anal.Appl.{\bf 10} 328.
\reff{22} Ratiu,T. 1980 J.Indiana.Math.Soc.{\bf 29} 609.
\reff{23} Hoppe,J. 1989 Int.J.Mod.Phys.{\bf A4} 5235.
\reff{24} Floratos,E.G. and Iliopoulos,J. 1988 Phys.Letts.{\bf B201} 237;
Floratos,E.G., Ilio-\break poulos,J. and Tiktopoulos,G. 1989 Phys.Letts.
{\bf B217} 285.
\reff{25} Yaffe,L.G. 1982 Rev.Mod.Phys.{\bf 54} 407.
\reff{26} Wootters,W.K. 1987 Annals of Phys.{\bf 176} 1.
\reff{27} Cohendet,O.,Combe,Ph.,Sirugue,M. and Sirugue-Collin,M. 1988 J.Phys.
{\bf A21} 2875.
\reff{28} Vourdas,A. 1990 Phys.Rev.{\bf A41} 1564,1653.
\reff{29} Groenewold,H.J. 1956 Kon.Dan.Vidensk.Selskab. {\bf 30} no.19.
\reff{30} Groenewold,H.J. 1946 Physica {\bf 12} 405.
\reff{31} Moyal,J. 1949 Proc.Camb.Phil.Soc.{\bf 45} 99.
\reff{32} Wigner,E.P. 1932 Phys.Rev.{\bf 40} 749.
\reff{33} Cohen,L. 1966 J.Math.Phys.{\bf 7} 781.
\reff{34} Argawal,G.S. and Wolf,E. 1970 Phys.Rev.{\bf D2} 2161,2187,2206.
\reff{35} Bakas,I. and Kakas,A.C. 1987 J.Phys.{\bf A20} 3713.
\reff{36} Baker,G.A. 1958 Phys.Rev.{\bf 109} 2198.
\reff{37} Wolski,A. and Dowker,J.S. 1991 J.Math.Phys.{\bf 32} 857,2304.
\reff{38} Epstein,P. 1903 Math.Ann. {\bf 56} 615.
\reff{39} Greenhill,A.G. 1878 Q.J.Math.{\bf 15} 15. See also 1892 {\it
Applications of elliptic functions} (Macmillan,London).
\reff{40} Hicks,W.M. 1878 Q.J.Math.{\bf 17}274.
\reff{41} Kronecker,L. 1889 Berliner Sitzungsberichte.
\reff{42} Oberhettinger,F. and Magnus,W. 1949 {\it Anwendung der Elliptische
Funktionen in Physik und Technik} (Springer-Verlag,Berlin).
\reff{43} The review article, Kraichnan,R. and Montgomery 1980 Reports on
Progress in Physics contains many references.
\reff{44} Kells,L.C. and Orszag,S.A. 1978 Phys.Fluids.{\bf 21} 162.
\reff{45} Orszag,S.A. 1971 Studies.Appl.Math.{\bf 50} 293.
\reff{46} Rankin,S.J. {\it {\rm SU(}$\infty${\rm )} and the large-$N$ limit}
DAMPT/91-30 Cambridge University.
\reff{47} Miller,J.,Weichman,P.B. and Cross,M.C. 1992 Phys.Rev.{\bf A45}.
\vfill\eject
\noindent FIGURE CAPTIONS.
\sect
\noindent Fig.1 Double vortex stream function for SU(9) plotted on the
torus at evolution times $t=$0, 1, 2, 3, 4 and 5.
\sect
\noindent Fig.2 Double vortex stream function for SU(31) plotted on the
torus at evolution times $t=$0, 0.5, 1.0 and 1.5.
\sect
\noindent Fig.3 Double vortex stream function for SU(5) plotted on the
torus at evolution times $t=$0, 1, 2, 3, 4 and 5.
\sect
\noindent Fig.4 SU(9) second moment of vorticity $M_2=<\ze_{\bf n}2>_t$, as
a function
of energy eigenvalue, $\La_{\bf n}$, calculated (by time averaging) at
several evolution times,
$t$. The diamonds correspond to the initial values, the crosses to $t=750$ and
the stars to $t=1500$. The circles indicate the values calculated on the basis
of a two-temperature canonical distribution. The initial configuration was the
two vortex one. The relaxation to the canonical values is better
than might have been expected in view of the relatively small number of
modes.
\sect
\noindent Fig.5 Second moment of vorticity and correlation function for three
modes of the SU(9) model as a function of time ($k2={\bf n}2$). The
horizontal levels indicate the value expected from a
canonical distribution. Note the start of the vertical axis. Although
the ordering of the levels is not reproduced, the general agreement is more
than acceptable.  The behaviour of the correlation function is consistent
with an ergodic development.
\sect
\noindent Fig.6 SU(9) long time evolution of the stream function starting from
the double vortex configuration. The evolution times are $t=$ 200, 400, 600 and
1000. See Fig.1 for the short time behaviour.
\end